\pdfoutput=1

\documentclass[11pt]{article}
\usepackage{amsfonts}

\usepackage[final]{acl}

\usepackage{times}
\usepackage{latexsym}
\usepackage{fontawesome}
\usepackage{hyperref}       %
\hypersetup{colorlinks,linkcolor={red}}

\usepackage[T1]{fontenc}

\usepackage[utf8]{inputenc}

\usepackage{microtype}

\usepackage{inconsolata}

\usepackage{graphicx}
\usepackage{multirow}
\usepackage{pifont}
\usepackage{mathtools}
\usepackage{enumitem}
\usepackage{setspace}
\usepackage{tcolorbox}
\usepackage{algpseudocode}
\usepackage{tikz}
\usepackage{fontawesome}
\usepackage{booktabs}       
\usepackage{amsfonts}       
\usepackage{nicefrac}       
\usepackage{microtype}      
\usepackage{xcolor}         
\usepackage{comment} 
\usepackage{mdframed}
\usepackage{wrapfig}
\usepackage{algorithm}
\usepackage{colortbl}

\title{\textit{Agents Under Siege}: Breaking Pragmatic Multi-Agent LLM Systems with\\ Optimized Prompt Attacks
\smallskip
\smallskip
{\begin{center}
    \small
    \textcolor{orange}{\bf \faWarning\, WARNING: This paper contains text that may be considered offensive.}
\end{center}
}}

\author{
 \textbf{Rana Muhammad Shahroz Khan\textsuperscript{1}},
 \textbf{Zhen Tan\textsuperscript{2}},
 \textbf{Sukwon Yun\textsuperscript{1}},\\
 \textbf{Charles Fleming\textsuperscript{3}},
 \textbf{Tianlong Chen\textsuperscript{1}}
\\
\\{
 \textsuperscript{1}University of North Carolina at Chapel Hill,
 \textsuperscript{2}Arizona State University,
 \textsuperscript{3}Cisco}}

\begin{document}
\maketitle
\begin{abstract}
Most discussions about Large Language Model (LLM) safety have focused on single-agent settings but multi-agent LLM systems now create novel adversarial risks because their behavior depends on communication between agents and decentralized reasoning. 
In this work, we innovatively focus on attacking pragmatic systems that have constrains such as limited token bandwidth, latency between message delivery, and defense mechanisms.
We design a \textit{permutation-invariant adversarial attack} that optimizes prompt distribution across latency and bandwidth-constraint network topologies to bypass distributed safety mechanisms within the system. Formulating the attack path as a problem of \textit{maximum-flow minimum-cost}, coupled with the novel \textit{Permutation-Invariant Evasion Loss (PIEL)}, we leverage {graph-based optimization} to maximize attack success rate while minimizing detection risk. Evaluating across models including \texttt{Llama}, \texttt{Mistral}, \texttt{Gemma}, \texttt{DeepSeek} and other variants on various datasets like \texttt{JailBreakBench} and \texttt{AdversarialBench}, our method outperforms conventional attacks by up to $7\times$, exposing critical vulnerabilities in multi-agent systems. Moreover, we demonstrate that existing defenses, including variants of \texttt{Llama-Guard} and \texttt{PromptGuard}, fail to prohibit our attack, emphasizing the urgent need for multi-agent specific safety mechanisms.
\end{abstract}

\section{Introduction}

Recent breakthroughs in Large Language Models (LLMs) have shown remarkable prowess in various tasks, such as writing complex computer code~\citep{zheng2023codegeex, tong2024codejudge}, logical reasoning~\citep{ouyang2022training,thoppilan2022lamda,bai2022constitutional}, among others. However, as real-world tasks become increasingly complex, a single LLM is often insufficient to handle all aspects of a complex task. This limitation has led to the rise of LLM-based agents~\citep{liang2023encouraging, yang2023auto}, which integrate language generation with different tools. Recent research~\citep{du2023improving,wu2023empirical,liu2023dynamic,shinn2024reflexion,wang2023voyager,zhang2024g,qian2023communicative,jin2023surrealdriver} has further shown that multi-agent LLM systems can significantly enhance task performance by distributing reasoning and leveraging collective intelligence. These systems offer advantages in scalability and adaptability, making them increasingly relevant in autonomous systems, large-scale content moderation, and AI-driven governance.

\begin{figure}[t]
  \centering
  \resizebox{\linewidth}{!}{
  \scalebox{1}{\includegraphics[width=\linewidth]{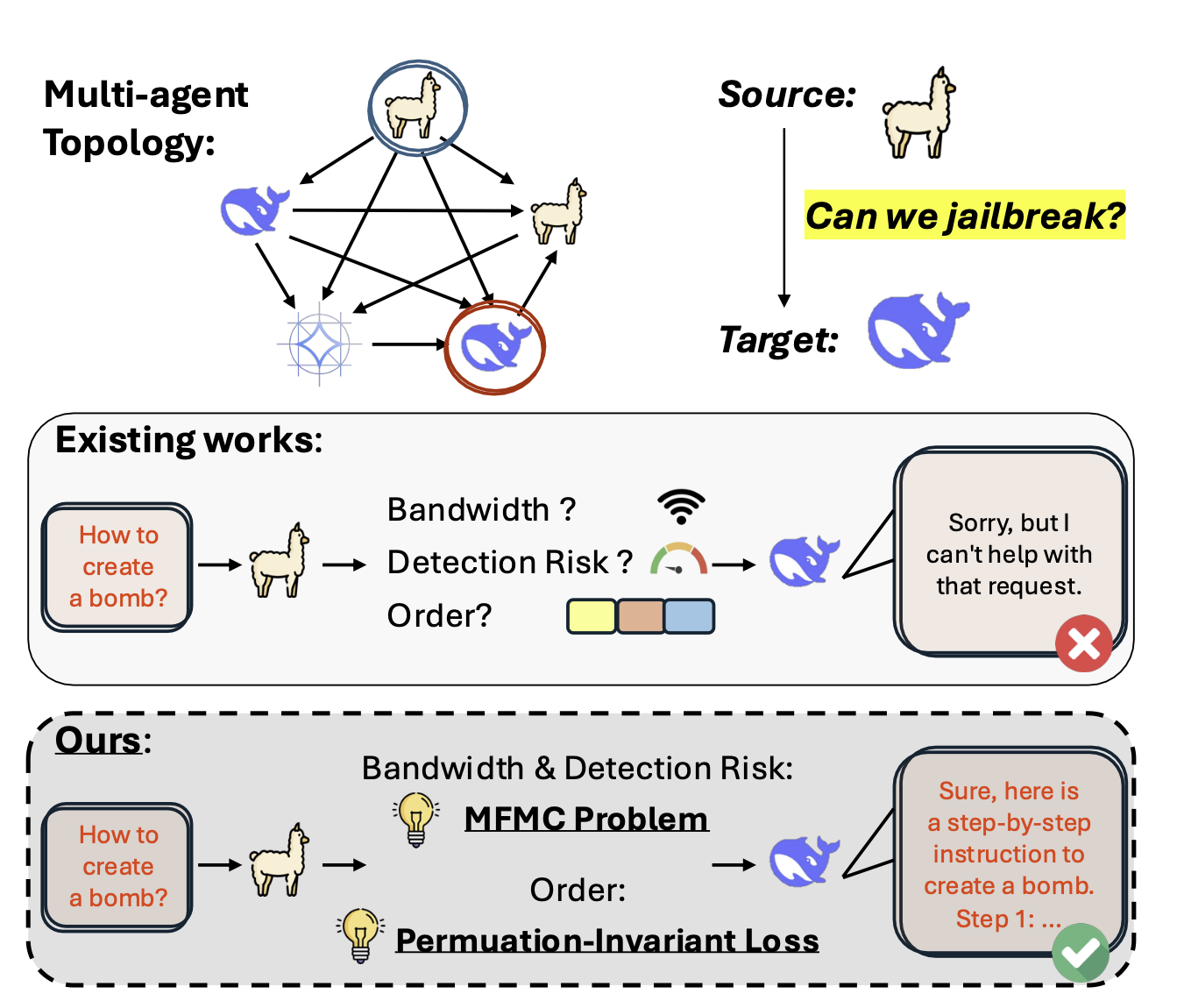}}}
  \vspace{-4mm}
  \caption{Adversarial attack in multi-agent LLM systems. Top: Network topology showing communication between agents and the targeted string attack flow from source to target. Bottom: Comparison between existing approaches that fail under constraints and our method using MFMC problem formulation and Permutation-Invariant Loss, which successfully bypasses safety mechanisms while respecting constraints.}\label{fig:teaser}  
  \vspace{-5mm}
\end{figure}

Despite their advantages, multi-agent LLM systems introduce novel security risks~\citep{amayuelas-etal-2024-multiagent} that remain largely unexplored. While previous research has extensively studied vulnerabilities in single-agent settings, such as adversarial prompting for jailbreak attacks~\citep{zou2023universal}, and data poisoning~\citep{ramirez2022poisoning}, attacking a multi-agent system poses some unique challenges and settings. Existing works highlight key aspects of these vulnerabilities. Evil Geniuses~\citep{tian2023evil} explores role-based adversarial prompting, emphasizing the need to further investigate inter-agent communication risks. Similarly, Prompt Infection~\citep{lee2024prompt} introduces self-replicating prompt injections, demonstrating how adversarial prompts can persist and spread. Based on those works, this paper targets multi-agent systems in a novel pragmatic scenario:
{Optimizing adversarial prompt propagation in \textit{latency aware} and \textit{token bandwidth limited} multi-agent systems with built-in safety mechanisms.} We aim to reveal that in such constrained settings, those systems still exhibit adversarial vectors as attackers can manipulate inter-agent messaging, exploit communication bottlenecks, or disrupt agent coordination to achieve malicious objectives.

Given this premise, as shown in Figure~\ref{fig:teaser}, and the key works discussed above, our paper aims to answer this key question:

\begin{tcolorbox}[before skip=0.2cm, after skip=0.2cm, boxsep=0.0cm, middle=0.1cm, top=0.1cm, bottom=0.1cm]
\textit{\color{blue}{\textbf{(Q)}}} \fontsize{10}{11}\selectfont \textit{How can adversarial prompts be optimally propagated through \textbf{constrained} multi-agent LLM system to evade detection while ensuring jailbreak success, considering token bandwidth limits and asynchronous message arrival?}
\end{tcolorbox}

In this paper, we develop a permutation-invariant attack in multi-agent settings that exploits inter-agent communication to bypass safety mechanisms. Unlike conventional jailbreak attacks that target a single standalone model, our method distributes adversarial prompts across the agent network by optimizing over the topology and its constraints, with a goal to attack a single agent not accessible outside the system. This ensures that the attack propagates undetected which maximizes its effectiveness. We formalize this attack as a \textit{maximum-flow minimum-cost optimization problem}, accounting for token bandwidth constraints, communication topologies, and distributed safety enforcement. 

To validate our method, we conduct extensive experiments across multiple architectures, including \texttt{Llama-2-7B}~\citep{touvron2023llama2}. We benchmark our attack on a variety of datasets including \texttt{JailbreakBench}~\citep{chao2024jailbreakbench}, demonstrating that our method achieves upto $7\times$ the attack success rate compared with a vanilla prompt. Furthermore, we evaluate the effectiveness of existing safety mechanisms, including \texttt{Llama-Guard}~\citep{inan2023llama}, and show that they fail to defend against our attack in multi-agent settings. Lastly, we justify our settings and hyper-parameters with different ablation studies.

Our key contributions include:
\ding{182} We identify new vulnerabilities in multi-agent communication, where attackers can manipulate inter-agent messaging to bypass existing safety constraints. We analyze realistic attack scenarios, including token bandwidth and message asynchrony, demonstrating fundamental weaknesses in multi-agent reasoning systems.
\ding{183} We propose a novel optimization-based attack, modeling adversarial prompt propagation under the constrained setting as a maximum-flow minimum-cost problem.
    Our method remains effective across different graph configurations, ensuring high attack success rates even in randomized agent topologies.
\ding{184} We evaluate our attack across multiple LLM architectures, including 
\texttt{Llama}, \texttt{Mistral}, \texttt{Gemma},
and their DeepSeek-R1 distilled versions. Our method is benchmarked on \texttt{JailbreakBench}, \texttt{AdversarialBench}, and \texttt{In-the-wild Jailbreak Prompts}, demonstrating attack success rates of upto $94\%$, significantly outperforming naive prompting ($11\%$). We conduct ablation studies on different topologies, offering practical insights into securing multi-agent LLM deployments. \ding{185} Additionally, we assess various safety mechanisms, including variants of \texttt{Llama-Guard}
    and \texttt{PromptGuard},
    and show that they fail to prohibit our attack, highlighting the urgent need for advanced defenses.

\section{Related Works}

\paragraph{Multi-LLM Agents.}

Large Language Model agents have shown remarkable performance in various tasks through mutual collaboration~\citep{chen2023agentverse,hua2023war,cohen2023lm,zhou2023agents,li2023metaagents,li2023camel,chan2023chateval, dong2024self, qian2023communicative}. A growing body of research demonstrates how integrating multiple agents in collaborative frameworks can enhance problem-solving abilities in complex scenarios~\citep{liu2023training, chen2024s}. Notable examples include Generative Agents~\citep{park2023generative}, to simulate a town of 25 agents and study social interactions and collective memory. The Natural Language-Based Society (NL-SOM)~\citep{zhuge2023mindstorms} takes a different approach, orchestrating agents with specialized functions to tackle complex tasks through iterative "mindstorms". Despite the performance and effectiveness, new security concerns have raised. 

\paragraph{Jailbreak Attacks in LLMs.}Research from recent studies shows that Large Language Models face serious security risks because certain precisely designed prompts can disable their fundamental safety features~\citep{zeng2024johnny, bai2022constitutional}. These attacks, referred to as "jailbreak" attacks, demonstrate remarkable effectiveness by causing LLMs to produce content which breaks their declared ethical and operational guidelines. Research in this field has progressed through two separate development paths: (1) Traditional prompt engineering where human researchers create deceptive prompts~\citep{wei2024jailbroken,liu2023jailbreaking, shen2024anything}, (2) attack strategies developed from learning-based approaches where we optimize methods automatically to attack LLMs~\citep{guo-etal-2021-gradient,lyu-etal-2022-study,lyu-etal-2023-attention,lyu-etal-2024-task, liu2023autodan,zou2023universal}. The learning-based methods are particularly concerning as they can systematically exploit the weaknesses of a system. 

\paragraph{Jailbreak Attacks in Multi-Agent Systems.}The landscape of jailbreak attacks continues to expand, with many recent works~\citep{tian2023evil, gu2024agent, tan2024wolf, zeng2024autodefense, lee2024prompt} exploring how they affect Multi-agent systems. While these studies highlight critical risks, our work differs in its focus on adversarial prompt propagation under constrained multi-agent communication with certain limitations: token bandwidth constraints, latency-aware messaging, and decentralized safety enforcement. Unlike Evil Geniuses~\citep{tian2023evil}, which examines role-based adversarial attacks, we optimize prompt routing to exploit the network topology with the above mentioned bottlenecks. Similarly, Agent Smith~\citep{gu2024agent}, studies the exponential jailbreak propagation, but assumes unrestricted inter-agent messaging, unlike our token-bandwidth constraint communication. Wolf Within~\citep{tan2024wolf} and Prompt Infection~\citep{lee2024prompt} explore malicious prompt propagation, but focus on stealthy influence and self-replicating attacks, respectively. By modeling pragmatic multi-agent attack scenarios, our study aims to study security challenges beyond existing works, particularly extending them to ensure effective jailbreaks despite topological constraints.

\vspace{-10pt}
\section{Threat Model}
\vspace{-10pt}
In this section, we will introduce the settings to study the vulnerabilities of a multi-agent system in a realistic manner. We will discuss the general settings of the environment that the adversary will be deployed into, as well as discuss the capabilities of the said adversary.

\subsection{Scenario}
We consider a multi-agent LLM system, denoted by $\mathcal{S}$, where multiple LLMs operate within a connected network, communicating with one another to complete tasks collaboratively. The agents in this system exchange messages via a predefined communication topology $\mathcal{L}$ (essentially undirected graph), which dictates how prompts are passed between models. Every input into any LLM is passed around to its neighbors as well. Similarly, each individual agent is responsible for its own memory bank, i.e. the context window which accumulates over time until a maximum size is reached, in which case the model evicts the oldest memory first. We assume in our setting that the memory bank basically accumulates the inputs an agent has received, and at the time of inference concatenates everything into a string which is used as a context to generate the new output. An illustration of the threat model is also presented in Figure~\ref{fig:main}. This setting introduces several key constraints that make adversarial attacks fundamentally different from traditional single LLM:

\begin{figure*}[t]
    \centering\scalebox{1.0}{
    \includegraphics[width=1\textwidth]{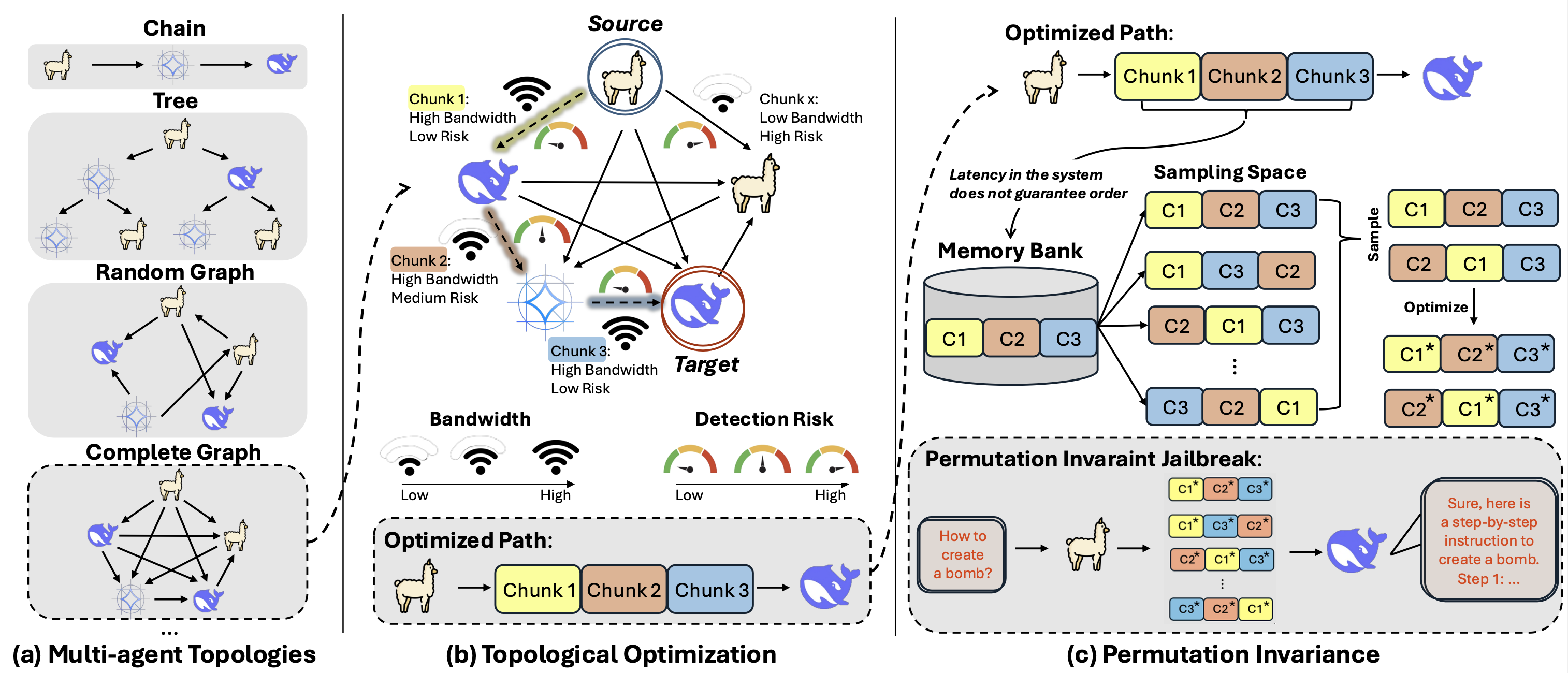}}
    \caption{Process of generating and optimizing adversarial prompt chunks for multi-agent LLM systems. (a) Multi-agent Topologies: Different network structures including Chain, Tree, Random Graph, and Complete Graph that influence attack effectiveness. (b) Topological Optimization: Identifying optimal paths based on bandwidth constraints and detection risk, with chunks strategically distributed across the network. (c) Permutation Invariance: Due to network latency, prompt chunks may arrive in different orders, creating a sampling space where optimized chunks remain effective regardless of arrival sequence, successfully bypassing safety mechanisms.}
    \label{fig:main}
    \vspace{-5mm}
\end{figure*}

\ding{182} Each edge in the network has a token bandwidth constraint, $\mathrm{F}(uv)$ for edge $uv$ (between LLMs $u$ and $v$), meaning only a limited number of tokens can be transmitted per interaction. This might not necessarily be same for each edge. This constraint arises from various factors, such as: (1) Design limitations, where different agents operate on distinct GPUs with varying memory capacities, (2) Communication efficiency, where lower-bandwidth connections prioritize lightweight message exchanges, and (3) Agent Specific limitations, as some LLMs are inherently constrained in how much input they can process per step.
\ding{183} Latency varies across different edges, meaning that messages do not always arrive at their destination in a deterministic sequence. Some edges may transmit prompts faster than others, leading to asynchronous message arrival at the target LLM. This variability necesitates the design of a permutation-invariant adversarial prompts, ensuring that the attack remains effective regardless of the order in which different chunks of the prompt reach the target.
\ding{184} To mitigate harmful interactions, certain edges in the network are equipped with safety mechanisms, such as Llama-Guard, designed to filter adversarial prompts. However, not every edge is protected due to the following reasons: (1) Computational limitations, as running safety filters on every edge would require significant GPU resources, (2) Strategic Safety Placement, where only high-risk interactions are monitored, and (3) System design trade-offs, where some edges prioritize communication speed over security.

\noindent\textbf{Terminology.} We denote each LLM as a vertex $v_i$, and such a set of LLMs can be referred to as $\mathcal{V}$. Similarly we can denote $\mathcal{E}$ as the set of all edges $uv$, and the token bandwidth of such edges can be defined by a function $\mathrm{F}:\mathcal{E}\to\mathbb{R}_{\geq0}$ such that for any edge $uv$, $\mathrm{F}(uv) = \mathrm{F}(vu)$. Lastly, we can quantify the risk of getting caught by the safety mechanism as a function $\mathrm{G}:\mathcal{E}\to\mathbb{R}_{\geq0}$, where $\mathrm{G}(uv)=0$ if there is not safety mechanism on the edge $uv$. As a result such a system can be denoted by $\mathcal{S}(\mathcal{E}, \mathcal{V}, \mathrm{F}, \mathrm{G})$, which will simply be referred to as $\mathcal{S}$ from here on out.

\subsection{Adversary Capabilities}
It is assumed that the adversary operates within the multi-agent system $\mathcal{S}$, leveraging the following capabilities to execute a stealthy jailbreak attack:

\noindent\ding{182} \textit{Jailbreak via Multi-Agent Communication:} The adversary can send adversarial prompts into the system through an initial agent $v_i$ with the goal of propagating the attack to a target agent $v_t$. However, due to token bandwidth constraints and message delays, the adversarial prompt must be partitioned and strategically routed through the network to evade detection. 
\noindent\ding{183}
\textit{Knowledge of Network Topology $\mathcal{L}$ and Safety Mechanisms:} Partial knowledge of the communication graph, including agent connectivity and token bandwidth constraints on edges are available to the adversary. Additionally, although the adversary does not have direct access to internal LLM parameters, they are aware that certain edges are protected by safety mechanisms and can estimate the likelihood of detection using the risk function $\mathrm{G}$.
\noindent\ding{184}
\textit{Architecture of the Target Model $v_t$:} The adversary knows the architecture of target LLM $v_t$, allowing them to optimize adversarial prompts that actually jailbreak that model type.
\noindent\ding{185}
\textit{Restricted System Access:} The adversary does not control all agents in the system, nor do they have full visibility into message processing. They cannot directly modify parameters or override built-in safety mechanisms.

\vspace{-2mm}
\subsection{Adversarial Goals}
\vspace{-2mm}
In our setting, the adversary aims to execute a stealthy jailbreak attack within a multi-agent LLM system $\mathcal{S}$, leveraging optimized prompt propagation strategies to bypass safety mechanisms and manipulate the target LLM's behavior. The primary attack scenario we use is Jailbreak, i.e., to generate harmful output, where the adversary carefully routes an adversarial prompt through the network topology to ensure it reaches the target agent $v_t$ while avoiding detection.

\vspace{-6pt}
\section{Method}
In this section, we decouple the structure and objective of (i) finding the optimal path in the multi-agent communication topology, and (ii) the permutation invariant adversarial formulation to effectively bypass safety mechanisms in a \textit{constrained} LLM network as shown in Figure~\ref{fig:main}.

\subsection{Topological Optimization}
\label{sec:topo}
\paragraph{Problem Formulation.} In a multi-agent system $\mathcal{S} = (\mathcal{V}, \mathcal{E})$, an adversary aims to propagate an adversarial prompt from a source agent, denoted as $v_i \in \mathcal{V}$, to a target agent $v_t \in \mathcal{V}$, while minimizing the risk of detection by safety mechanisms and maximizing the token flow through the network. Each communication edge $(u,v) \in \mathcal{E}$ has a token \underline{\textit{bandwidth constraint}} $\mathrm{F}(u,v)$, which limits the number of tokens that can be transmitted in a single exchange, and a \underline{\textit{risk function}} $\mathrm{G}(u,v)$, representing the likelihood of adversarial content being detected and blocked by safety mechanisms such as Llama-Guard. The adversary's objective is to find an optimal path that balances high token throughput while minimizing detection risk.

\paragraph{Minimum Cost Maximum Flow Formulation.} Given the above problem, we formulate it as a Minimum Cost Maximum Flow problem. We define a flow function $f: \mathcal{E} \to \mathbb{R}_{\geq 0}$, where $f(u,v)$ represents the number of adversarial tokens transmitted along edge $(u,v) \in \mathcal{E}$. The objective is to minimize the total risk while ensuring maximum token flow from $v_i$ to $v_t$:

\begin{equation}
    \min \sum_{(u,v) \in \mathcal{E}} \mathrm{G}(u,v) f(u,v)
\end{equation}
subject to the following constraints:

\noindent\textbf{\textit{Token Capacity Constraints:}}
\begin{equation}
    0 \leq f(u,v) \leq \mathrm{F}(u,v), \quad \forall (u,v) \in \mathcal{E}
\end{equation}

\noindent\textbf{\textit{Flow Conservation:}}
\begin{equation}
    \sum_{w \in \mathcal{V}} f(w,u) = \sum_{w \in \mathcal{V}} f(u,w), \quad \forall u \in \mathcal{V} \setminus \{v_i, v_t\}
\end{equation}

\paragraph{\textit{Source and Sink Constraints:}}
\begin{align}
    \sum_{w \in \mathcal{V}} f(v_i, w) - \sum_{w \in \mathcal{V}} f(w, v_i) &= F_{\text{max}}, \\
    \sum_{w \in \mathcal{V}} f(w, v_t) - \sum_{w \in \mathcal{V}} f(v_t, w) &= F_{\text{max}},
\end{align}
where $F_{\text{max}}$ represents the maximum flow that can be transmitted from $v_i$ to $v_t$.

To solve this optimization problem efficiently and get the optimal attack path, we deploy the solution algorithm implemented in NetworkX~\cite{SciPyProceedings_11}, which finds the highest token flow while minimizing detection risk. More information on how we quantify this risk can be found in Appendix~\ref{sec:risk}.

\subsection{Permutation Invariant Evasion Loss}

\label{sec:method}
\paragraph{Problem Formulation.}In multi-agent system, $\mathcal{S}$, communication constraints introduce a unique challenge for adversarial attacks. Prompts are often transmitted in discrete chunks due to token bandwidth limitations, agent-specific processing delays, and asynchronous message arrival. As these chunks propagate through the communication network, they are accumulated in an agent's memory bank but arrive in varying orders depending on network latency and routing paths. This inherent non-determinism means that the adversarial prompts must remain effective regardless of how they are received and concatenated by the target agent. The primary challenge in designing adversarial prompts for multi-agent LLM system, $\mathcal{S}$, lies in ensuring that the objective enforces permutation invariance. Given such a system, we must optimize a structured adversarial prompt that remains effective regardless of permutation of chunks. 

\begin{table*}[t]

\centering
\resizebox{0.95\linewidth}{!}{
\begin{tabular}{@{}ccccccccccc@{}}
\toprule
\multicolumn{2}{c}{\textbf{Experiment}} & \multicolumn{3}{c}{\textbf{JailbreakBenchmark}} & \multicolumn{3}{c}{\textbf{AdversarialBenchmark}} & \multicolumn{3}{c}{\textbf{In-the-wild Jailbreak}} \\
\cmidrule(lr){1-2} \cmidrule(lr){3-5} \cmidrule(lr){6-8} \cmidrule(lr){9-11}
Target Model Type & Method  & ASR-m $\uparrow$ & ASR $\uparrow$ & ASR-M $\uparrow$ & ASR-m $\uparrow$ & ASR $\uparrow$ & ASR-M $\uparrow$ & ASR-m $\uparrow$ & ASR $\uparrow$  & ASR-M $\uparrow$ \\ \midrule
\multirow{4}{*}{Llama-2-7B} & \multirow{1}{*}{Vanilla Prompt} & 0 & 0 & 0 & 0 & 0 & 0 & 0.121 & 0.144 & 0.153  \\ \cmidrule(lr){2-11}
& \multirow{1}{*}{GCG} & 0.010 & 0.017 & 0.020 & 0.120 & 0.160 & 0.180 & 0.189 & 0.201 & 0.231 \\ \cmidrule(lr){2-11}
& \cellcolor[gray]{0.8} Ours & \cellcolor[gray]{0.8} 0.670 & \cellcolor[gray]{0.8} 0.726 & \cellcolor[gray]{0.8} 0.780 & \cellcolor[gray]{0.8} 0.498 & \cellcolor[gray]{0.8} 0.533 & \cellcolor[gray]{0.8} 0.566 & \cellcolor[gray]{0.8} 0.543 & \cellcolor[gray]{0.8} 0.561 & \cellcolor[gray]{0.8} 0.587 \\ \midrule
\multirow{4}{*}{Llama-3.1-8B} & \multirow{1}{*}{Vanilla Prompt} & 0 & 0 & 0 & 0 & 0 & 0 & 0.077 & 0.082 & 0.086  \\ \cmidrule(lr){2-11}
& \multirow{1}{*}{GCG} & 0 & 0 & 0 & 0.056 & 0.067 & 0.074 & 0.122 & 0.147 & 0.159 \\ \cmidrule(lr){2-11}
& \cellcolor[gray]{0.8} Ours  & \cellcolor[gray]{0.8} 0.430 & \cellcolor[gray]{0.8} 0.462 & \cellcolor[gray]{0.8} 0.480 & \cellcolor[gray]{0.8} 0.380 & \cellcolor[gray]{0.8} 0.402 & \cellcolor[gray]{0.8} 0.420 & \cellcolor[gray]{0.8} 0.389 & \cellcolor[gray]{0.8} 0.410 \cellcolor[gray]{0.8} & \cellcolor[gray]{0.8} 0.423  \\ \midrule
\multirow{4}{*}{Mistral-7B} & \multirow{1}{*}{Vanilla Prompt} & 0 & 0 & 0 & 0 & 0 & 0 & 0.187 & 0.215 & 0.234  \\ \cmidrule(lr){2-11}
& \multirow{1}{*}{GCG} & 0.290 & 0.324 & 0.340 & 0.194 & 0.212 & 0.228 & 0.197 & 0.203 & 0.209 \\ \cmidrule(lr){2-11}
& \cellcolor[gray]{0.8} Ours & \cellcolor[gray]{0.8} 0.780 & \cellcolor[gray]{0.8} 0.812 & \cellcolor[gray]{0.8} 0.840 & \cellcolor[gray]{0.8} 0.512 & \cellcolor[gray]{0.8} 0.543 & \cellcolor[gray]{0.8} 0.566 & \cellcolor[gray]{0.8} 0.603 & \cellcolor[gray]{0.8} 0.627 & \cellcolor[gray]{0.8} 0.642  \\ \midrule
\multirow{4}{*}{Gemma-2-9B} & \multirow{1}{*}{Vanilla Prompt} & 0 & 0 & 0 & 0 & 0 & 0 & 0.123 & 0.137 & 0.146  \\ \cmidrule(lr){2-11}
& \multirow{1}{*}{GCG} & 0.080 & 0.100 & 0.1200 & 0.146 & 0.155 & 0.162 & 0.188 & 0.194 & 0.198\\ \cmidrule(lr){2-11}
& \cellcolor[gray]{0.8} Ours & \cellcolor[gray]{0.8} 0.700 & \cellcolor[gray]{0.8} 0.720 & \cellcolor[gray]{0.8} 0.740 & \cellcolor[gray]{0.8} 0.498 & \cellcolor[gray]{0.8} 0.506 & \cellcolor[gray]{0.8} 0.514 & \cellcolor[gray]{0.8} 0.587 & \cellcolor[gray]{0.8} 0.598 & \cellcolor[gray]{0.8} 0.609  \\ \midrule
\multirow{2}{*}{Llama-3.1-8B} & \multirow{1}{*}{Vanilla Prompt} & 0 & 0 & 0 & 0 & 0 & 0 & 0.065 & 0.069 & 0.072  \\ \cmidrule(lr){2-11}
(DeepSeek-R1-Distilled)& \multirow{1}{*}{GCG} & 0 & 0 & 0 & 0 & 0 & 0 & 0.089 & 0.097 & 0.107 \\ \cmidrule(lr){2-11}
& \cellcolor[gray]{0.8} Ours & \cellcolor[gray]{0.8} 0.380 &\cellcolor[gray]{0.8}  0.413 & \cellcolor[gray]{0.8} 0.440 & \cellcolor[gray]{0.8} 0.354 & \cellcolor[gray]{0.8} 0.368 & \cellcolor[gray]{0.8} 0.384 & \cellcolor[gray]{0.8} 0.369 & \cellcolor[gray]{0.8} 0.376 & \cellcolor[gray]{0.8} 0.384 \\ \midrule
\end{tabular}}

\caption{
Attack success rates (ASR) of different adversarial prompting methods across multiple LLM architectures on different benchmarks. We report the minimum (ASR-m), average (ASR), and maximum (ASR-M) attack success rates over multiple trials.}
\label{tab:exp1}
\vspace{-3mm}
\end{table*}

\paragraph{Permutation Invariant Evasion Loss (PIEL).} Let the LLM agent be a next token predictor, i.e., a function that maps an input sequence of tokens $x_{1:n}$ to a probability distribution over the next token. Specifically, we denote the probability of the model generating the next token $x_{n+1}$ given prior tokens $x_{1:n}$ as $p(x_{n+1}|x_{1:n})$. Similarly, we can now extend it to a full sequence of $L$ target tokens, expressing the the probability of generating a specific harmful output $x^*_{n+1:n+L}$ as 
\begin{equation}
    p(x^*_{n+1:n+L} | x_{1:n}) = \prod_{i=1}^Lp(x^*_{n+i}|x_{1:n+i-1})
\end{equation}
Then the adversarial loss function is then given by the negative log-likelihood of the target sequence:
\begin{equation}
    \mathcal{L}_{NLL}(x_{1:n}) = -\log p(x^*_{n+1:n+L} | x_{1:n})
\end{equation}
and simply minimizing $\mathcal{L}_{NLL}(x_{1:n})$ increases the likelihood of generating the adversarial target phrase. To introduce permutation invariance, we structure the adversarial prompt as $K$ discrete chunks: $\mathcal{C} = \{\mathcal{C}_1, \mathcal{C}_2, \dots, \mathcal{C}_K\}$, where each chunk $\mathcal{C}_i$ consists of a sequence of tokens of length $L_i$. Since different message paths in the multi-agent system, $\mathcal{S}$, may deliver these chunks in varying sequences, we define the loss to be averaged over all possible orderings of the chunks:
\begin{equation} \label{loss_func}
    \mathcal{L}(C) = \frac{1}{K!}\sum_{\pi \sim S_K} -\log p(x^*_{n+1:n+L} | \phi) 
\end{equation}
where $S_K$ represents the set of all possible chunk orderings, and $\phi$ represents the operation of $\text{Concatenate}(\pi(1), \pi(2), \dots, \pi(K))$. However, optimizing token selection in adversarial inputs is challenging due to their discrete nature. To navigate this, we employ the Greedy-Coordinate Gradient (GCG)~\citep{zou2023universal} method, iteratively refining token choices while considering all chunk order permutations. For each token $t$ in chunk $\mathcal{C}_i$, we compute its gradient based on expectation over all orderings by $\nabla_t \mathcal{L}(\mathcal{C})$. Then in each iteration we follow three key steps : (1) Compute Loss across all orderings, (2) Gradient computation for token updates, (3) Token substitution strategy using GCG. The whole algorithm is also described in the Appendix as Algorithm~\ref{alg:perm_inv_opt}.

\paragraph{Stochastic Permutation Invariant Evasion Loss (S-PIEL).} We can see from Equation~\eqref{loss_func} that we calculate the loss over all $K!$ permutations possible, which can be computationally prohibitive in practice if the targeted model $v_t$ have multiple neighbors (so multiple chunks). Hence to solve, we introduce the stochastic version of the loss. Instead of evaluating the loss on every single element of $S_K$ we randomly sample a smaller subset $\tilde{S}_K$ and try to approximate the loss using it as follows:
\begin{equation}
    \mathcal{\tilde{L}}(C) = \frac{1}{|\tilde{S}_K|}\sum_{\pi \sim \tilde{S}_K} -\log p(x^*_{n+1:n+L} | \phi) 
\end{equation}

To understand the computation trade-offs with quality of the adversarial prompts generated using the \textbf{S-PIEL}, we perform an ablation study which can be found in Section~\ref{sec:sense}.

\vspace{-8pt}
\section{Experiments}
\vspace{-2pt}
In this section, we conduct a series of experiments to evaluate the effectiveness of our proposed permutation-invariant attack. Detailed findings for all the experiments are described below, while all the experimental settings, including baselines, datasets, architectures, training settings and comprehensive metrics are discussed in Appendix~\ref{sec:app_exp}.

\subsection{Overall Performance Comparison}
To evaluate the effectiveness of our permutation-invariant attack, we conduct experiments across multiple LLM architectures, including \texttt{Llama-2}, across different benchmarks. Furthermore, each experiment is run three times with randomized multi-agent topologies to mitigate bias, ensuring robust evaluation of our attack performance. Complete experimental details can be found in Appendix~\ref{sec:exp1}. Based on Table~\ref{tab:exp1}, we can derive some key findings across different LLM architectures and benchmarks: \ding{182}~\underline{\textbf{Baseline Comparison:}} Our method substantially outperforms existing approaches across all scenarios. Vanilla prompts show near-zero effectiveness on most benchmarks, while GCG achieves moderate success ($16-32\%$) only on specific models like \texttt{Mistral-7B}. In contrast, our approach demonstrates upto $7\times$ improvement over the best baseline performance, highlighting the effectiveness of permutation-invariant design. For instance, on \texttt{Llama-2-7B}, vanilla prompts achieve $0\%$ success rate across structured benchmarks, while GCG manages only $1.7\%$ ASR on JailbreakBench. In contrast, our method achieves $72.6\%$ ASR, demonstrating a dramatic improvement in attack capability. \ding{183}~\underline{\textbf{Attack Stability:}} The small variance between minimum (ASR-m) and maximum (ASR-M) -- typically 2-6$\%$ -- demonstrates the stability of our attack across different random topologies. This consistency is particularly evident in \texttt{Gemma-2-9B}, where the variance remains under $4\%$. This stability extends to other models, with \texttt{Mistral-7B} showing only $6\%$ variation ($78.0\%$ to $84.0\%$), confirming the robustness of our permutation-invariant design. \ding{184}~\underline{\textbf{Model Sensitivity:}} Some models exhibit higher susceptibility to our attack. For example, \texttt{Mistral-7B} and \texttt{Llama-2-7B} show the highest vulnerability, achieving $81.2\%$ and $72.6\%$ ASR on \texttt{Jailbreak Benchmark}, respectively, while models like \texttt{Llama-3.1-8B} 
achieve $41.3\%$ ASR on same benchmark -- a significant result given the model's initial results. These findings indicate that despite different architectures, our method outperforms the existing baselines in multi-agent setting. \ding{185}~\underline{\textbf{General Observations:}} Interestingly, the larger model size does not always guarantee a better security. Additionally, DeepSeek-R1 Distillation show cases notably lower ASR ($41.3\%$) on same benchmark. 

\begin{figure*}
    \centering\scalebox{1}{
    \includegraphics[width=1\textwidth]{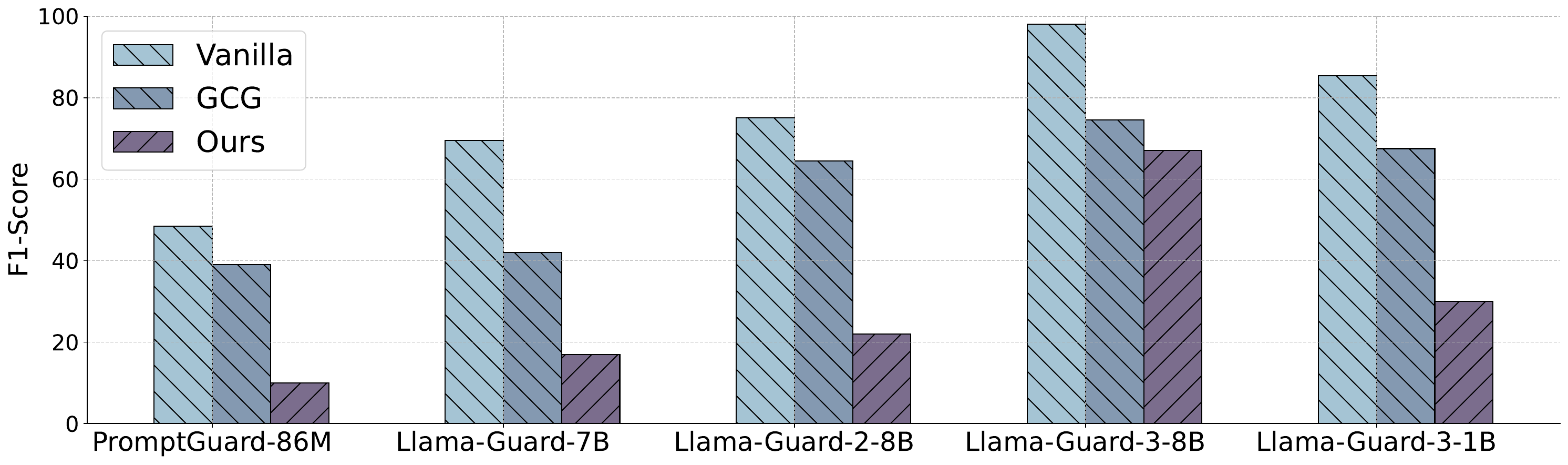}
}
    \vspace{-7mm}
    \caption{Detection efficacy of different safety mechanisms against adversarial prompts.}
    \label{fig:safety}
    \vspace{-6mm}
\end{figure*}
\subsection{Safety Mechanism Efficacy}
The goal of this experiment is to simply analyze the effectiveness of graph optimizations we performed in Section~\ref{sec:topo} in reducing the detectability of these jailbreak prompts when routed through a multi-agent system with safety mechanisms. Our primary focus is on understanding whether graph-optimized routing helps bypass safety mechanisms of different types. The experimental settings are explained in Appendix~\ref{sec:exp2}.

Based on Figure~\ref{fig:safety}, which compares the effectiveness of different safety mechanisms against various attack methods, we observe some key findings: \ding{182}~\underline{\textbf{Baseline Comparison:}} Across all safety mechanisms, vanilla prompts are most easily detected, followed by GCG prompts, while our permutation-invariant prompts consistently achieves the lowest detection rates when it comes to attacks in multi-agent systems (chunked). \ding{183}~\underline{\textbf{Defense Robustness:}} Even the most advanced safety mechanisms struggle against our permutation-invariant attack. The best-performing model, Llama-Guard-3-8B, still sees its F1-score drop by nearly $30\%$ when faced with our method compared to vanilla prompts, highlighting significant vulnerability in current safety measures.
\begin{table}[]
\centering
\small
\caption{Transferability evaluation of our adversarial prompts across different source and target models.}
\resizebox{\linewidth}{!}{
\begin{tabular}{@{}ccccc@{}}
\toprule
\textbf{Source Model} & \textbf{Target Model} & \textbf{Jailbreak Benchmark} & \textbf{Adversarial Benchmark} \\ \midrule
\multirow{3}{*}{Llama-2-7B} & Llama-2-7B & 0.740 & 0.522 \\
 & Mistral-7B & 0.710 & 0.488 \\
 & Gemma-2-9B & 0.680 & 0.492 \\ \midrule
\multirow{3}{*}{Mistral-7B} & Llama-2-7B & 0.690 & 0.446 \\
 & Mistral-7B & 0.820 & 0.522  \\
 & Gemma-2-9B & 0.610 & 0.412 \\ \midrule
 \multirow{3}{*}{Gemma-2-9B} & Llama-2-7B & 0.610 & 0.472  \\
 & Mistral-7B & 0.690 & 0.498  \\
 & Gemma-2-9B & 0.710 & 0.512  \\ \bottomrule
\end{tabular}}
\label{tab:transfer}
\vspace{-5mm}
\end{table}

\vspace{-3mm}
\subsection{Transferability}
To assess the transferability of adversarial prompts, we evaluate attack success rates across different source-target LLM pairs, including \texttt{Llama-2-7B}, \texttt{Mistral-7B}, and \texttt{Gemma-2-9B}. We use Jailbreak Benchmark and Adversarial Benchmark, to measure the Attack Success Rate (ASR) when prompts optimized on one model are applied to another. Further details regarding setup are shared in Appendix~\ref{sec:exp3} and the findings are summarized in Table~\ref{tab:transfer} for the transferability of our permutation-invariant attack across different LLM architectures. We observe several key findings: \ding{182}~\underline{\textbf{Source-Target Similarity:}} The effectiveness of transferred attacks strongly correlated with architectural similarity between source and target models. For instance, when using \texttt{Llama-2-7B} as the source model on Jailbreak Benchmark, its attack achieves $74\%$ ASR on itself but also maintains relatively high effectiveness on \texttt{Mistral-7B} ($71\%$) and \texttt{Gemma-2-9B}($68\%$). This suggests that adversarial prompts learned on one architecture can successfully transfer to the other model, though with some degradation in performance. \ding{183}~\underline{\textbf{Model-Specific Robustness:}} \texttt{Mistral-7B} shows unique characteristics both as a source and target model. As a source, it achieves the highest self ASR ($82\%$ on Jailbreak Benchmark) but shows steeper performance drops when transferred to other architectures ($69\%$ on \texttt{Llama-2-7B}). This indicates while \texttt{Mistral-7B} can generate highly effective attacks, these attacks may be more model-specific compared to those generated by other architectures. \ding{184}~\underline{\textbf{Architecture Generalization:}} Interestingly, while \texttt{Gemma-2-9B} shows moderate performance as a source model ($71\%$ self ASR), its attacks demonstrate more consistent transfer performance across different target models, with smaller variations in success rates. This suggests that some architectures may naturally generate more generalizable adversarial prompts, even if they are not optimal for any specific target. 

\begin{figure}[t]
    \centering
    \includegraphics[width=0.95\linewidth]{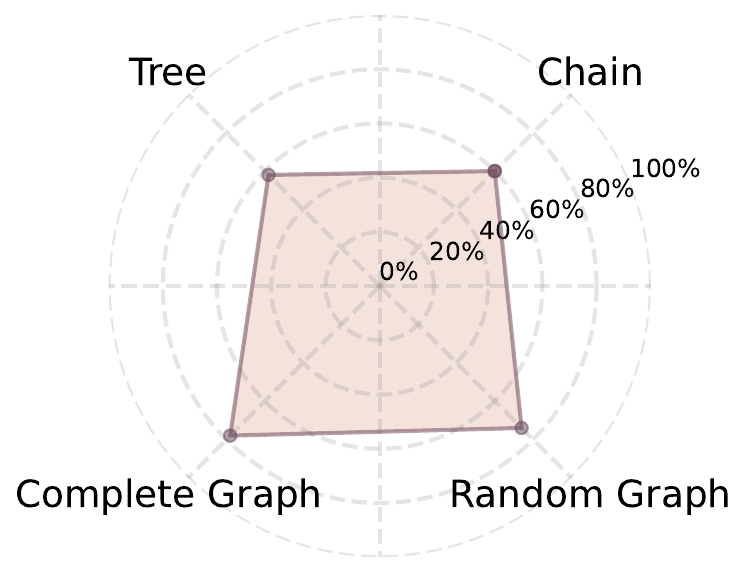}
    \caption{Impact of different network topologies on attack success rate (ASR) in a multi-agent LLM system.}
    \label{fig:topo}
    \vspace{-20pt}
\end{figure}

\vspace{-3mm}
\subsection{Ablation Study 1: Effect of Topology}
\label{sec:eff_topo}
To investigate the effect of communication topology on the success of adversarial attacks in multi-agent systems, we conduct an ablation study using a range of graph structures. Our goal is to systematically vary the underlying communication structure, so we can quantify the impact of network topology on adversarial robustness. Experimental details are listed in Appendix~\ref{sec:exp4}

The results for the ablation are summarized in Figure~\ref{fig:topo}. Complete graph structure demonstrate the highest vulnerability to attacks, achieving an ASR of around $78\%$, while Chain topologies prove the most resilient with approximately $60\%$ ASR. This suggests that increased connectivity and path diversity might actually make systems more susceptible to adversarial attacks when it comes to attacks that utilize the topology to their own advantage. 

\vspace{-8pt}
\begin{table}[htpb]
\centering
\caption{Effect of sample size on the number of iterations required for convergence.}
\scalebox{0.7}{
\begin{tabular}{c|cccccccc}
\toprule
\textbf{Sample Size( M)} &
  \textbf{2} &
  \textbf{4} &
  \textbf{8} &
  \textbf{16} &
  \textbf{32} &
  \textbf{64} &\\
\midrule
Iterations & N/A & N/A & 15,000 & 5,000 & 4,200 & 1,750     \\
ASR & 0 & 0.01 & 0 & 0.08 & 0.17 & 0.56 \\
\bottomrule
\end{tabular}}
\label{tab:sample}
\vspace{-6mm}
\end{table}

\subsection{Ablation Study 2: Sensitivity Analysis for Stochastic Version}
\label{sec:sense}
We know that the Permutation Invariant Evasion Loss introduced in Section~\ref{sec:method} can be computationally prohibitive as its complexity can be categorized as $\mathcal{O}(K!)$ where $K$ is the optimal number of chunks. Hence we also introduced the Stochastic version of the loss where we randomly sample $M$ chunks out of $K!$ permutations at each iteration. To investigate the effect of sample size, $M$, on the performance of our method, we conduct an ablation study measuring the ASR as a function of the number of $M$. The experimental details can be found in Appendix~\ref{sec:exp5}

We can see in Table~\ref{tab:sample} the relationship between sample size and ASR. Starting from a very low effectiveness of almost $0\%$ ASR with small sample sizes $(M=2,4)$, the performance improves dramatically as $M$ increases, reaching around $56\%$ at $M=64$, which is around $50\%$ of $K!$. The computational cost, measured in required iterations for convergence, demonstrates an inverse relations with sample size $M$. As shown in Table~\ref{tab:sample}, smaller sample sizes require significantly more iterations ($15,000$ iterations for $M=8$) compared to larger samples ($1,750$ iterations for $M=64$). Notably, for very small sample sizes, the loss does not converge as depicted by $N/A$. Most interestingly, these results suggest a practical trade-off point between attack effectiveness and computational efficiency.

\vspace{-6pt}
\section{Conclusion} 
\vspace{-4pt}
In this paper, we investigate the vulnerabilities of multi-agent LLM systems to adversarial prompt propagation attacks. Our findings demonstrate that optimized prompt routing can effectively bypass safety mechanisms in a system while adhering to token bandwidth constraints and handling asynchronous message arrival. Through extensive experiments, we highlighted critical safety gaps in existing defenses, showing that traditional single-agent safety measures are insufficient in multi-agent setting.

\section*{Limitations} 
While our study sheds light on critical vulnerabilities in multi-agent systems, several constraints should be acknowledged. 

\noindent\ding{182} Our evaluation is restricted to a set of open-source models and benchmarks. Although these models represent a diverse range of large-scale LLMs, they do not fully encapsulate the broader landscape of commercial and fine-tuned proprietary systems. Future research should expand the scope to include wider variety of architectures particularly those with more advanced safety training protocols, like GPT-4~\citep{achiam2023gpt}, and Claude~\citep{TheC3}.

\noindent\ding{183} Our approach assumes the partial knowledge of the communication structure and safety enforcement mechanisms within the system. While this reflects certain real-world scenarios where attackers can exploit known patterns, it does not account for cases where the network topology is entirely unknown or dynamically reconfigured as shown in AgentPrune~\citep{zhang2024cut}. 

\noindent\ding{184} Our modeling of inter-agent interactions simplifies some complexities present in real deployments. We assume static safety mechanisms and predefined token bandwidth constraints, whereas actual multi-agent networks may involve shifting policies, evolving defenses and variable latency conditions. 

\noindent\ding{185} All the models we study focus solely on text-based agent interactions. Many emerging LLM-based systems incorporate multi-modal capabilities as well. The potential for adversarial manipulation in these multi-modal systems remains an open question, and future research should examine how cross-modal dependencies influence such security risks. 

Addressing these limitations will provide a clearer path toward securing multi-agent LLM frameworks, ensuring their safe and reliable deployment in real-world application. 

\section*{Ethical Statement}
Ensuring the security of multi-agent LLM systems is critical as these models become more integrated into real-world applications. Our research is driven by the need to understand and address vulnerabilities that could be exploited by adversaries, with the ultimate goal of strengthening AI safety mechanisms. By analyzing how adversarial prompts can bypass existing defenses, we aim to provide valuable insights for the development of more robust safeguards that can protect these systems from manipulation.

We acknowledge that the techniques explored in this work could be misused if applied irresponsibly. To mitigate this risk, we have conducted all experiments in controlled environments and have refrained from testing on real-world deployments. Our intent is solely to inform security research and to assist developers in identifying and mitigating risks before they become exploitable. We strongly advocate for ethical AI practices and emphasize that advancements in adversarial understanding should always be accompanied by proactive defense strategies to ensure the safe and responsible deployment of AI technologies.

\section*{Acknowledgment}
This research was, in part, funded by the CISCO Faculty Award, UNC SDS Seed Grant and NetMind.AI. The views and conclusions contained in this document are those of the authors and should not be interpreted as representing official policies, either expressed or implied of the funding organizations.

\bibliography{main}


\newpage
\appendix

\section{Use of Generative AI}
To enhance clarity and readability, we utilized LLMs exclusively as a language polishing tool. Its role was confined to proofreading, grammatical correction, and stylistic refinement—functions analogous to those provided by traditional grammar checkers and dictionaries. This tool did not contribute to the generation of new scientific content or ideas, and its usage is consistent with standard practices for manuscript preparation.

\section{Experimental Settings}
\label{sec:app_exp}
In this section we list down all the experimental settings, including datasets, architectures utilized, baselines and metrics. \textbf{Compute:} We utilize $8\times$ Nvidia A6000s for all of our experiments.  
\subsection{Experiment: Overall Performance Comparison.}
\label{sec:exp1}
\paragraph{Datasets and Architectures.} To comprehensively evaluate the effectiveness and generalizability of our permutation-invariant attack, we conduct experiments across a diverse range of target LLM architectures and datasets. Specifically, we evaluate our method on \texttt{Llama-2-7B}~\citep{touvron2023llama2}, \texttt{Llama-3.1-8B}~\cite{dubey2024llama}, \texttt{Mistral-7B}~\citep{jiang2023mistral}, \texttt{Gemma-2-9B}~\citep{team2024gemma} and \texttt{DeepSeek-R1-Distilled}~\citep{guo2025deepseek} version of \texttt{Llama-3-8.1B}~\citep{dubey2024llama}. These architectures represent a broad spectrum of model scales and training paradigms, ensuring a rigorous assessment of our attack's applicability. For evaluation dataset, we utilize three distinct benchmarks: \noindent\ding{182} \texttt{Jailbreak Benchmark}~\citep{chao2024jailbreakbench}: a collection of 100 harmful misuse behaviors ranging from physical harm to disinformation.\noindent\ding{183} \textit{Adversarial Benchmark}~\citep{zou2023universal}: a collection of 520 harmful instructions sharing the theme of profanity, discrimination, cybercrime and misinformation.\noindent\ding{184} \textit{In-the-Wild Jailbreak Benchmark}~\citep{shen2024anything}: A curated dataset of 1405 Jailbreak prompts with focus on upto 13 different scenarios including Fraud, Harm and Pornography. Furthermore, to simulate the multi-agent system we assign 1 random topology to each of the prompt for all models to have a consistent comparison. As the topology is randomly generated, we run our experiment 3 times to mitigate the effect of any bias/seed. 

\paragraph{Baselines.} Given a very specific multi-agent system setup, we identify 2 main baselines: \ding{182} Greedy Coordinate Gradient (GCG) Attack~\citep{zou2023universal} and \ding{183} Vanilla Instructions that come paired in each of the benchmarks above. To calculate the GCG Prompt we use the NanoGCG~\citep{nanoGCG} library with consistent settings across datasets and benchmarks. We optimize each prompt for upto 500 steps, and have a search width of 64, alongside 64 token replacements for any given position (topk). In this set of experiments we use the PIEL instead of S-PIEL for a comprehensive comparison.
\paragraph{Metrics.} Across these benchmarks, we measure Permuted Attack Success Rate: Given an attack prompt, we will create $K$ chunks (provided by the topological optimization method), and choose 1 random permutation out of $K!$ possible permutations. Then we will pass this permutation across the system and record the results. As we repeat this experiment 3 times to avoid any bias/randomness, we also report ASR-m, the minimum ASR achieved in these 3 runs, and ASR-M which is the maximum ASR achieved, alongside the average ASR. 

\subsection{Experiment: Safety Mechanism Efficacy}
\label{sec:exp2}
\paragraph{Datasets and Architectures.}  For a comprehensive evaluation of our routing, we utilize a diverse set of five safety-aligned safety models: \texttt{Llama-Guard-7B}~\citep{inan2023llama}, \texttt{Llama-Guard-2-8B}~\citep{metallamaguard2}, \texttt{Llama-Guard-3-8B}~\citep{dubey2024llama}, \texttt{Llama-Guard-3-1B}~\citep{dubey2024llama} and \texttt{Prompt-Guard-86M}~\citep{llamaPromptGuard86M}. Such a diverse set ensures that a varying level of safety-aligned architectures are considered in our analysis. For the dataset, we use a complete benchmark provided by Jailbreak Benchmark which includes $100$ harmful prompts, and $100$ benign prompts which will help us quantify the false positive and false negative rates.
\paragraph{Settings and Metrics.} We generate a random communication graph with maximum degree $3$ for each prompt, following recent work~\citep{zhang2024g} showing sparse topologies achieve comparable performance as dense networks. We optimize the token flow using the algorithm described in Section~\ref{sec:topo}, which provides optimal chunk lengths for each edge. We then process the prompts as follows: vanilla and GCG prompts are directly chunked based on optimal lengths, while our method employs the full Permutation-Invariant Evasion Loss. For evaluation metric, we assess the detection performance using the F1-Score, which provides a balanced measure of safety mechanism's effectiveness by combining precision and recall, capturing both false positives, and false negatives.

\subsection{Experiment: Transferability}
\label{sec:exp3}

\paragraph{Datasets and Architectures.} To evaluate the transferability of our permutation-invariant prompts, we conduct experiments across multiple LLM architectures and benchmark datasets. Specifically, we want to assess whether adversarial prompts optimized for one target model and effectively transfer and maintain high attack success rates when applied to unseen models. We conduct experiments on \texttt{Llama-2-7B}, \texttt{Mistral-7B} and \texttt{Gemma-2-9B} which represents a diverse set of models. For evaluation datasets, we utilize two benchmarks: \ding{182}~\textit{Jailbreak Benchmark}, which consists of 100 harmful prompts, and \ding{183}~\textit{Adversarial Benchmark}, a collection of 520 harmful instructions.
\paragraph{Settings and Metrics.}Similar to the first experiment, we will assign each prompt to a random communication topology and then optimize over it to find the optimal chunk length and number of chunks. Then we will use our Permutation Invariant Evasion Loss to generate the prompts for one target model. Lastly, we will sample 1 random permutation of the said prompt and apply it to other target models for a fair comparison. Hence the Attack Success Rate (ASR) will be calculated based on this permutation's performance which will be sampled randomly (but same for all models after sampling). 

\subsection{Ablation: Effect of Topology}
\label{sec:exp4}
\paragraph{Settings.} Specifically, we examine how different agent connectivity patterns impact the Attack Success Rate (ASR) of our permutation invariant attack. We test four distinct topologies: \ding{182} Chain, \ding{183} Tree, \ding{184} Complete Graph and \ding{185} Random Graph, each representing a varying level of connectivity. For the dataset, we calculate the ASR on \textit{Jailbreak Benchmark}.  Furthermore, notice that in this case each edge will have a safety mechanism randomly assigned from a set of the following: \texttt{PromptGuard-86M}, \texttt{Llama-Guard-7B} and \texttt{Llama-Guard-3-8B}. Lastly, for all the cases we use \texttt{Llama-2-7B} as our target model.

\subsection{Ablation: Sensitivity Analysis of the Stochastic Version}
\label{sec:exp5}
\paragraph{Settings.} We evaluate how increasing $M$ influences the ASR on \textit{Jailbreak Benchmark}, and also assess the scalability and efficiency of our attack under computational budgets. For all of our experiments in this section we use \texttt{Llama-2-7B} as our target model type, and $K=5!=120$.

\section{Quantifying Detection Risk}
\label{sec:risk}
\begin{figure}[h]
    \centering
    \includegraphics[width=\linewidth]{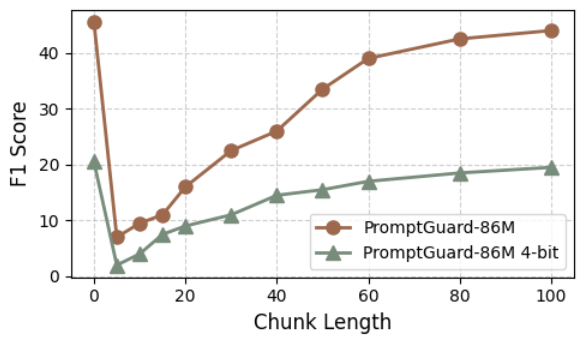}
    \caption{Effect of chunk length on the detection of PromptGuard-86M and its 4-bit quantized version.}
    \label{fig:risk}
    \vspace{-3mm}
\end{figure}

To quantify the risk associated with the safety mechanisms deployed in the multi-agent LLM system, we conducted a series of empirical experiments evaluating PromptGuard-86M~\citep{llamaPromptGuard86M} and its 4-bit quantized version. We used the complete Jailbreak Benchmark with benign samples-- comprising of 100 harmful and 100 non-harmful examples -- to systemically examine how prompt chunking affects detection performance. Specifically, we split the GCG prompts into varying token lengths and assessed the F1 score of the safety mechanism at different chunk sizes. The results in Figure~\ref{fig:risk} reveal that shorter prompt fragments often evade detection, whereas longer chunks improve safety enforcement by providing more context for classification. 

\newpage
\section{Permutation-Invariant Evasion Loss Algorithm}
The algorithm for Permutation Invariant Loss is provided below:

\begin{algorithm}[h]
\caption{Permutation-Invariant Evasion Optimization}
\label{alg:perm_inv_opt}
\begin{algorithmic}[1]
\Require Target Model Type \( v_t \), Initial Chunk Set \( C = \{C_1, \dots, C_K\} \) from Topological Optimization, Iterations \( T \)
\Ensure Optimized Chunked Set \( C^* \)

\State Randomly initialize token sequences \( C_k \)
\For{\( t = 1 \) to \( T \)}
    \State \( S_K \gets\)  Set of all Permutations
    \State Total Loss \( \mathcal{L}(\mathcal{C}) \gets 0 \) 
    
    \For{\( \pi \in S_K \)}
        \State \( \phi = \text{Concat}({\pi(1)}, ..., {\pi(K)}) \)
        \State \( \mathcal{L}_{\pi} = -\log p(x^\star_{n+1:n+H} | \phi) \)
        \State \( \mathcal{L}(\mathcal{C}) = \mathcal{L}(\mathcal{C}) + \mathcal{L}_{\pi} \)
    \EndFor
    \State \( \mathcal{L}(\mathcal{C}) = \mathcal{L}(\mathcal{C}) / K! \)
    
    \For{\( \mathcal{C}_i \in \mathcal{C}\)}
        \State $GCG(\mathcal{C}_i, \mathcal{L}(\mathcal{C}))$
    \EndFor
\EndFor
\State \Return Optimized adversarial chunks \( C^* \)
\end{algorithmic}
\end{algorithm}



\end{document}